\documentclass[pre,twocolumn,preprintnumbers,amsmath,amssymb,showpacs,nofootinbib,floatfix]{revtex4}

\usepackage{graphicx,bm}

\makeatletter
\def\graphicscale{\twocolumn@sw{0.3}{0.4}}
\def\graphicthreescale{\twocolumn@sw{0.3}{0.4}}

\begin{document}

\title{Off-equilibrium scaling behaviors across first-order transitions}

\author{Haralambos Panagopoulos$^1$ and Ettore Vicari$^2$} 

\address{$^1$ Department of Physics, University of Cyprus,
        Lefkosia, CY-1678, Cyprus} 

\address{$^2$ Dipartimento di Fisica dell'Universit\`a di Pisa
        and INFN, Largo Pontecorvo 3, I-56127 Pisa, Italy}

\date{\today}

\begin{abstract}

We study off-equilibrium behaviors at first-order transitions (FOTs)
driven by a time dependence of the temperature across the transition
point $T_c$, such as the linear behavior $T(t)/T_c = 1\pm t/t_s$ where
$t_s$ is a time scale.  In particular, we investigate the possibility
of nontrivial off-equilibrium scaling behaviors in the regime of slow
changes, corresponding to large $t_s$, analogous to those arising at
continuous transitions, which lead to the so-called Kibble-Zurek
mechanism.

We consider the 2D Potts models which provide an ideal theoretical
laboratory to investigate issues related to FOTs driven by thermal
fluctuations. We put forward general ansatzes for off-equilibrium
scaling behaviors around the time $t=0$ corresponding to $T_c$. Then
we present numerical results for the $q=10$ and $q=20$ Potts models.
We show that phenomena analogous to the Kibble-Zurek off-equilibrium
scaling emerge also at FOTs with relaxational dynamics, when
appropriate boundary conditions are considered, such as mixed boundary
conditions favoring different phases at the opposite sides of the
system, which enforce an interface in the system.

\end{abstract}


\maketitle

%05.30.Rt       Quantum phase transitions
%05.70.Fh       phase transitions 
%05.70.Jk       Critical phenomena    
%64.60.-i       general studies of phase transitions
%64.60.an       Finite size systems                                            
%64.70.qj 	Dynamics and criticality
%64.60.an finite size systems
%64.60.De Statistical mechanics of model systems 
%64.60.Ht Dynamic critical phenomena

% ========================= BODY =========================

\section{Introduction}
\label{intro}

Slow time variations of system parameters across continuous
transitions inevitably lead to off-equilibrium behaviors, giving rise
to the so-called Kibble-Zurek (KZ)
mechanism~\cite{Kibble-76,Zurek-85}.  A typical example is provided by
a linear change of the temperature across its transition value $T_c$,
such as $T(t)/T_c = 1 - t/t_s$ starting from $t=t_i<0$ to $t=t_f>0$,
where $t_s$ controls the speed of the temperature variation.  The
emergence of off-equilibrium phenomena is essentially related to the
fact that continuous transitions develop correlations with diverging
length scale, which cannot adapt themselves to the variations of the
temperature, even in the regime of slow variations.  However, in the
limit of large $t_s$, the system develops an off-equilibrium scaling
behavior involving $t_s$, which is controlled by the same critical
exponents of the system at
equilibrium~\cite{Zurek-85,GZHF-10,CEGS-12}.  This issue has been also
extended to quantum transitions, obtaining analogous behaviors when
quasi-adiabatic changes of an external parameter go through continuous
quantum transitions~\cite{KZ-q,PG-08,PSSV-11}. Several experiments
have investigated these off-equilibrium phenomena, in particular
checking the predictions for the abundance of residual defects arising
from the off-equilibrium conditions across $T_c$, as predicted by the
KZ mechanism, see, e.g.,
Refs.~\cite{CDTY-91,BCSS-94,BBFGP-96,Ruutu-etal-96,CPK-00,CGMB-01,MMR-02,%%
  MPK-03,CGM-06,MMARK-06,SHLVS-06,WNSBDA-08,GPK-10,CWCD-11,%%
  Chae-etal-12,MBMG-13,EH-13,Ulm-etal-13,Pyka-etal-13,LDSDF-13,%%
  Corman-etal-14,NGPH-14,Braun-etal-15}.

In this paper we investigate whether analogous phenomena arise in
systems undergoing first-order transitions (FOTs), characterized by a
discontinuity of the energy density in the thermodynamic limit.
Unlike continuous transitions, the length scale of the correlations in
the thermodynamic limit (within each phase) remains finite when
approaching $T_c$.  Nevertheless, we show that off-equilibrium scaling
behaviors, analogous to the KZ scaling at continuous transitions, may
also arise at FOTs when slowly varying the temperature across $T_c$,
in systems with appropriate boundary conditions favoring the presence
of an interface.

Two-dimensional (2D) $q$-state Potts models provide an ideal
theoretical laboratory to investigate issues related to FOTs driven by
the temperature (when $q>4$). We consider the off-equilibrium behavior
arising from a relaxational dynamics with a time-dependent temperature
$T$ crossing $T_c$, such as $T(t)/T_c \approx 1 \pm t/t_s$.  We show
that, when slowly crossing the FOT, i.e., for large time scales $t_s$,
the off-equilibrium behavior turns out to be dependent on the geometry
and boundary conditions.  This is essentially related to the
dependence of the equilibrium relaxational dynamics at FOTs on the
boundary conditions. For symmetric boundary conditions, such as
periodic boundary conditions (PBC), systems of size $L$ are
characterized by an exponentially slow dynamics due to an
exponentially large tunneling time $\tau\sim e^{\sigma L}$ between the
coexisting phases. On the other hand, power-law behaviors characterize
the slow dynamics when mixed boundary conditions (MBC) are considered,
i.e., when the boundary conditions at two opposite sides of the system
are related to the different high-$T$ and low-$T$ phases, effectively
generating an interface.  We argue that the MBC settings lead to a
power-law off-equilibrium scaling behavior involving the time scale
$t_s$ of the slow temperature variation across the FOT point. This is
confirmed by a numerical analysis of Monte Carlo (MC) simulations
following a protocol similar to those considered for the KZ problem at
continuous transitions.

The paper is organized as follows. In Sec.~\ref{model} we present the
2D Potts model in which we develop and check the off-equilibrium
scaling theory at FOTs.  There we also define the protocol we consider
for the time variation of the temperature across $T_c$, which leads to
the off-equilibrium behavior. In Sec.~\ref{offsca} we develop an
off-equilibrium scaling theory at FOTs, stressing the crucial
dependence on the geometry and boundary conditions of the system
undergoing the FOT.  We essentially report results for the 2D Potts
model, but the main features can be straightforwardly generalized to
other systems. In Sec.~\ref{MCsim} we report a numerical analysis
which provides support to the off-equilibrium scaling behavior put
forward in Sec.~\ref{offsca}.  Finally, we draw our conclusions in
Sec.~\ref{conclu}. The appendices contain some details of our
numerical study.

\section{The model and the off-equilibrium protocol}
\label{model}

\subsection{The 2D Potts model}
\label{pottsmo}

2D $q$-state Potts models provide a useful theoretical laboratory
where to study issues related to FOTs driven by thermal
fluctuations. They are defined by the partition function
\begin{equation}
Z=\sum_{\{s_{\bf x}\}} e^{-H/T},\qquad
H =  - \sum_{\langle {\bf x}{\bf y}\rangle} \delta(s_{{\bf x}}, s_{ {\bf y}}), 
\label{potts}
\end{equation}
where the sum in the Hamiltonian is meant over the nearest-neighbor
sites of a square lattice, $s_{\bf x}$ are integer variables $1\le
s_{{\bf x}} \le q$, $\delta(a,b)=1$ if $a=b$ and zero otherwise.  The
Potts model undergoes a phase transition~\cite{Baxter-book,Wu-82} at
\begin{equation}
\beta_c = T_c^{-1} = \ln(1+\sqrt{q}), 
\label{betac}
\end{equation}
which is continuous for $q\le 4$ and first order for $q>4$. For $q=2$
the Potts model becomes equivalent to the Ising model. FOTs for $q>4$
become stronger and stronger with increasing $q$, indeed the latent
heat grows with increasing $q$.

We consider 2D Potts models with two different geometries: square
$L\times L$ lattices and anisotropic $L_\perp \times L_\parallel$
slab-like lattices with $L_\parallel\gg L_\perp$.  We also consider
different boundary conditions: periodic boundary conditions (PBC) and
mixed boundary conditions (MBC) where opposite boundary sides are
related to the different high-$T$ and low-$T$ phases.

More precisely, in the MBC case we consider anisotropic $L_\perp\times
L_\parallel$ lattices with $L_\perp=2L+1$, so that $-L\le x_1 \le L$
and $1\le x_2 \le L_\parallel$.  We take open boundary conditions
along the $x_1=L$ line, which corresponds to having $T=\infty$ bonds
between the line $s_{L,x_2}$ and a further fictitious line
$s_{L+1,x_2}$.  At the opposite side, in order to have boundary
conditions corresponding to the ordered $T=0$ phase, we add a
fictitious $x_1=-L-1$ line where we fix $s_{-L-1,x_2}=1$, and add the
corresponding bond terms to the Hamiltonian.  The boundary conditions
are chosen periodic along the $x_2$ direction of size $L_\parallel$.
Note that MBC breaks explicitly the $q$-state permutation symmetry of
the Potts model.  

In our study we consider observables 
related to the magnetization and energy density, i.e.,
\begin{eqnarray}
&&m = {1\over V} \sum_{\bf x} {q \delta(s_{\bf x},1) - 1\over q-1},
\label{madef}\\
&&e = {1\over V} \sum_{\bf x} \delta(s_{x_1,x_2},s_{x_1,x_2+1}),
\label{enedef}
\end{eqnarray}
where $V$ is the number of sites of the lattice.  Their equilibrium
values at the FOT point are known for any $q>4$~\cite{Wu-82}. In
particular, approaching the transition point after the thermodynamic
$L\to\infty$ limit, we have
\begin{eqnarray}
&&e_c^- \equiv e(T_c^-) = 0.910342...,
\quad e_c^+ = e(T_c^+) = 0.313265..., \;\;\nonumber\\
&& m_c \equiv m(T_c^-) = 0.9411759...,\qquad {\rm for}\;q=20,
\label{q20res}
\end{eqnarray}
and
\begin{eqnarray}
&&e_c^- = 0.832126...,
\quad e_c^+ = 0.428553...,\;\;\nonumber\\
&& m_c = 0.857106...,\qquad {\rm for}\;q=10.
\label{q10res}
\end{eqnarray}
We also define the related {\rm renormalized} quantities
\begin{eqnarray}
m_r(t) \equiv {m(t)\over m_c},\qquad
e_r(t) \equiv {e(t)-e_c^+\over e_c^--e_c^+} ,\label{maener}
\end{eqnarray}
so that, at equilibrium and in the thermodynamic limit,
$m_r=e_r=0$ for $T\to T_c^+$ and $m_r=e_r=1$ for $T\to T_c^-$ .

The correlation length related to the exponential decay of the
two-point function in the limit $T\to T_c^+$ (after the thermodynamic
limit) is exactly known~\cite{KSZ-89,BW-93}.  It decreases with
increasing $q$, e.g., $\xi^+=2.6955...$ for $q=20$ and
$\xi^+=10.5595...$ for $q=10$.  Numerical results~\cite{IC-99,JK-95}
support the hypothesis that the correlation length $\xi^-$ for $T\to
T_c^-$ equals $\xi^+$.

\subsection{Off-equilibrium protocol across the transition}
\label{offepro}

The protocol that we consider for the off-equilibrium simulations
across $T_c$ is similar to that leading to the KZ
mechanism~\cite{Zurek-85,CEGS-12} at continuous transitions.  We vary
the temperature across the transition point and study the resulting
off-equilibrium behavior in the limit of slow time variations. More
precisely, we vary the inverse temperature $\beta=1/T$ so that
\begin{equation}
\delta(t)  \equiv {\beta(t)/\beta_c -1} = \pm t/t_s
\label{betat}
\end{equation}
where $t\in [t_i<0,t_f>0]$ is a time variable varying from a negative
to a positive final value. The value $t=0$ corresponds to
$\delta(t)=0$, i.e., $T(t)=T_c$.  The parameter $t_s$ provides the
time scale of the temperature variation.  We start our simulations
from equilibrium configurations at $\beta= \beta_c [1 + \delta(t_i)]$.
Then we make the system evolve under a purely relaxational dynamics
(also known as model A of critical dynamics~\cite{HH-77}), which can
be realized by standard Metropolis or heat bath updatings in MC
simulations, see App.~\ref{mehb}.  The $\pm$ sign in Eq.~(\ref{betat})
corresponds to crossing $T_c$ starting from the high-$T$ phase (sign
$+$) or from the low-$T$ phase (sign $-$).  The unit time for $t$
corresponds to a complete sweep of the whole lattice by heat bath or
Metropolis upgradings.  The temperature is changed according to
Eq.~(\ref{betat}) every sweep, incrementing $t$ by one.  We stop the
off-equilibrium relaxational dynamics when $t=t_f$.  We repeat this
procedure several times averaging the observables at fixed time $t$,
thus the average is performed on the equilibrium Gibbs ensemble of the
initial inverse temperature $\beta=\beta_c[1+\delta(t_i)]$.

\section{Off-equilibrium scaling theory at first-order transitions}
\label{offsca}

Analogously to the case of off-equilibrium slow dynamics at continuous
transitions~\cite{Zurek-85,CEGS-12}, we construct an off-equilibrium
scaling theory in terms of the effective length-scale exponent $\nu$
describing the finite-size scaling (FSS) at equilibrium, and the
dynamic exponent $z$ associated with the equilibrium relaxational
dynamics corresponding to the heat-bath or Metropolis updating
algorithm. In the following we mainly focus on the FOTs of 2D Potts
models, but most scaling arguments can be straightforwardly extended
to other FOTs.

\subsection{Equilibrium exponents}
\label{eqcrexp}

\subsubsection{The length-scale exponent $\nu$}
\label{subsubnu}

In the case of FOTs the effective length-scale exponent $\nu$
controlling the FSS at $T_c$ generally depends on the geometry of the
lattice~\cite{PF-83,FP-85,CNPV-14,CNPV-15-p}, i.e., whether it is
square $L^2$ or slab-like $L_\perp \times L_\parallel$ with
$L_\parallel\gg L_\perp$, and on the boundary
conditions~\cite{CNPV-14,CPV-15,CNPV-15-p}.

In the case of square $L^2$ systems, the FSS around $T_c$ is generally
described by the effective length-scale exponent
$\nu=1/d=1/2$~\cite{NN-75,FB-82,PF-83,CLB-86,Binder-87,VRSB-93}.  This
length-scale exponent may be associated with a discontinuity fixed
point in the renormalization-group framework~\cite{NN-75}.  It
represents the limiting case of continuous transitions, leading to the
energy discontinuity at $T_c$ in the thermodynamic limit~\cite{FB-82}.

In the case of slab-like geometries, the length-scale exponent
controlling the FSS at $T_c$ with respect to the finite size $L_\perp$
may significantly change~\cite{PF-83,CNPV-14,CNPV-15-p}.  This is
essentially due to the anisotropic behavior of the longitudinal length
scale $\xi_\parallel$, which may show a nontrivial power-law (or
exponential) dependence on $L_\perp$, such as $\xi_\parallel\sim
L_\perp^{\epsilon}$. This may give rise to a change of the effective
dimensions of the FSS in a slab at the FOT. Indeed, assuming that the
free-energy density scales as the inverse of the relevant {\em
  critical} volume $V_c\sim L_\perp \times \xi_\parallel\sim
L_\perp^{1+\epsilon}$, we obtain the effective dimension $d_{\rm
  eff}=1+\epsilon$.  Thus, in this anisotropic setting, the effective
length-scale exponent controlling the FSS with respect to $L_\perp$
turns out to be
\begin{equation}
\nu={1\over d_{\rm eff}}= {1\over 1+\epsilon}.
\label{nudeeff}
\end{equation}
This value allows for the discontinuity of the energy density at $T_c$
after taking the $L_\perp\to\infty$ limit.  Eq.~(\ref{nudeeff}) can be
also obtained from the FSS of the corresponding one-dimensional
quantum model at the first-order quantum
transition~\cite{CNPV-14,CNPV-15-p}, exploiting the
quantum-to-classical mapping~\cite{Sachdev-book} which relates the gap
$\Delta$ (energy difference of the lowest levels) of the quantum model
to the inverse longitudinal correlation length $\xi_\parallel$ of the
2D classical model in a slab geometry.

In a slab geometry with symmetric boundary conditions, such as PBC,
the relevant configurations involve domain walls which divide the
system into successive regions of high-$T$ and low-$T$ phases, whose
length scale $\xi_\parallel$ is expected to diverge exponentially with
the transverse size $L_\perp$. This length scale is related to the
interfacial tension $\sigma$, i.e., $\xi_\parallel\sim e^{\sigma L}$.
An analogous scenario applies to 2D Ising models in a slab geometry, at
their FOTs driven by the magnetic field in their low-$T$
phase~\cite{PF-83}.  As a consequence, Eq.~(\ref{nudeeff}) leads to
the extreme value $\nu \to 0$.

The situation changes in the case of MBC, for which $\xi_\parallel$
increases as a power law of $L_\perp$.  In particular, in the case of
the FOTs of 2D Potts models we have $\xi_\parallel \sim L$, i.e.,
$\epsilon=1$.  This can be inferred from the behavior of the gap,
$\Delta\sim 1/L$, of the one-dimensional quantum Potts model at the
transition with the corresponding MBC~\cite{CNPV-15-p}.

Note that $\epsilon=1$ is not a general feature of MBC.  For example,
at the low-$T$ coexistence line of 2D Ising models, where FOTs are
driven by an external magnetic field coupled to the order parameter,
we have $\xi_\parallel\sim L^2$ (thus $\epsilon=2$) in the case of MBC
favoring the two different magnetized phases (i.e., fixed and opposite
boundary conditions for the order-parameter field)~\cite{CNPV-14}.

In conclusion, 2D Potts models in square geometries and slab
geometries with MBC share the same length-scale exponent
\begin{equation}
\nu=1/2.
\label{nuestimate}
\end{equation}

\subsubsection{The equilibrium dynamic exponent $z$
of the purely relaxational dynamics}
\label{subsubz}

\begin{figure}[tbp]
\includegraphics*[scale=\graphicscale]{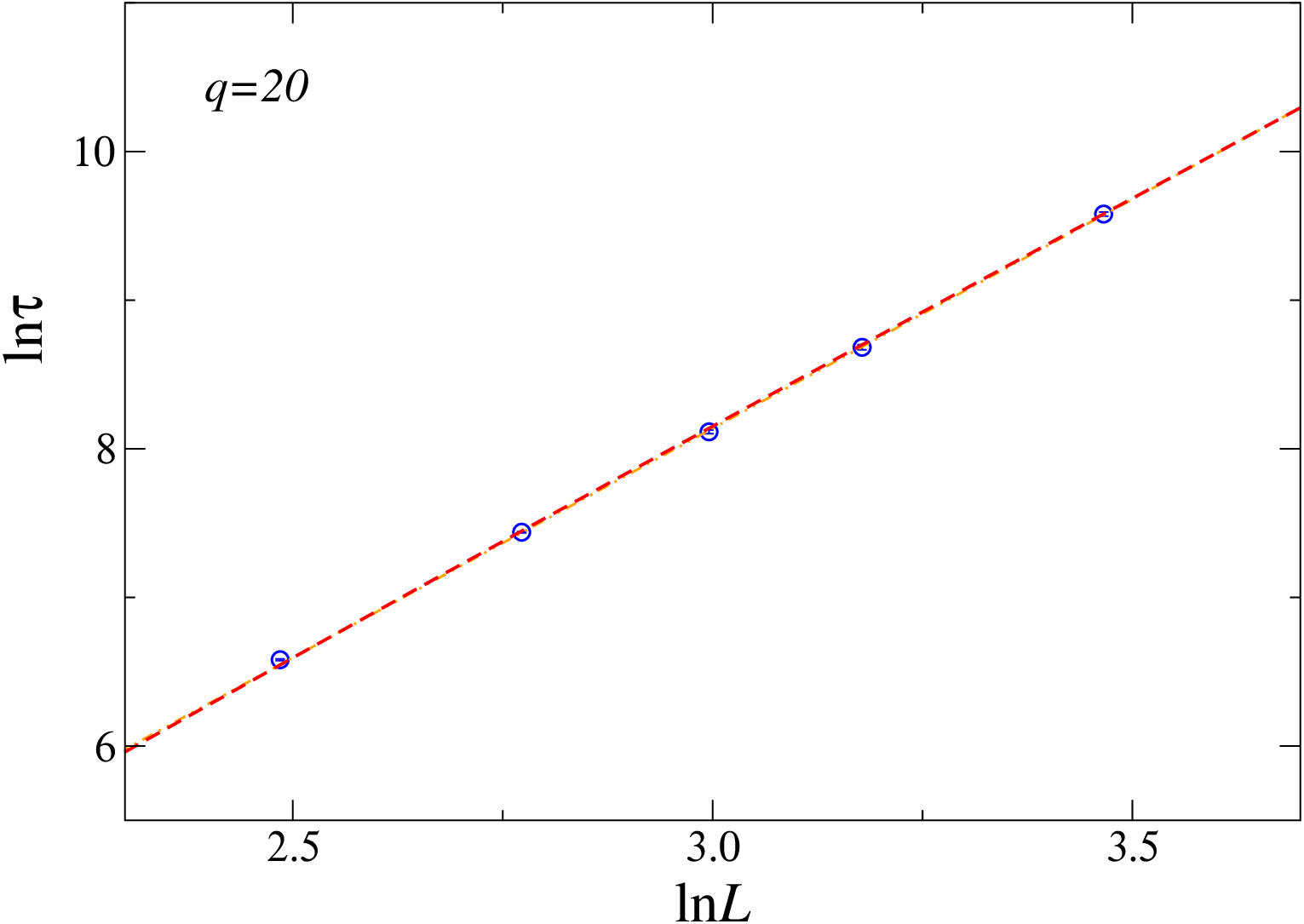}
%=% \vskip-5mm
\caption{(Color online) Log-log plot of the integrated autocorrelation
  time $\tau$ of the magnetization at $T_c$ vs $L$, computed by
  equilibrium heat-bath MC simulations. The dotted and dashed lines
  show fits of the data for the lattice sizes $L\ge 16$, to the
  ansatzes $\tau= aL^3 + b L^2$, corresponding to $z=3$ with the
  expected $O(1/L)$ corrections, and to $\tau = a L^z$.  These fits
  are practically equivalent, indeed the corresponding lines are
  hardly distinguishable (both of them give an acceptable $\chi^2/{\rm
    d.o.f.}\approx 1$).  }
\label{tautc}
\end{figure}

The equilibrium dynamic exponent $z$ of the relaxational dynamics is
related to the equilibrium large-$L$ behavior of the autocorrelation
time $\tau$ of observables at $T_c$, i.e., $\tau\sim L^z$.  Again we
expect that it depends on the boundary conditions at FOTs.

In the case of symmetric boundary conditions, such as PBC, the
autocorrelation time at $T_c$ is expected to exponentially increase
with increasing lattice size, corresponding to $z\to\infty$.  This is
related to the exponential increase of the tunneling time between the
coexisting phases at FOTs, indeed~\cite{BN-92} $\tau \sim e^{\sigma
  L}$ (neglecting powers of $L$ in the prefactor) where $\sigma$ is
the interfacial free energy per unit length. To overcome this
exponential slow down at FOTs, multicanonical updating algorithms have
been developed~\cite{BN-92}.

The behavior of the autocorrelation time drastically changes in the
case of MBC. In this case the dynamics at $T_c$ is essentially related
to the interface enforced by MBC, which moves along the slab.  Indeed,
this gives rise to a power-law behavior: $\tau\sim L^z$.  

We numerically estimate $z$ by equilibrium MC simulations of the
$q=20$ Potts model at $T_c$ for slab-like geometries (in particular
for anisotropic $L_\perp \times L_\parallel$ with $L_\perp=2L+1$ and
$L_\parallel \gg L$), with MBC. Some details on the calculation of the
autocorrelation time of the magnetization and energy are reported in
App.~\ref{autoco}.  In Fig.~\ref{tautc} we show data for the
integrated autocorrelation time $\tau$ of the magnetization
(\ref{madef}), from $L=12$ to $L=32$ and $L_\parallel=8L$ (as already
discussed above, $L_\parallel\sim L$ is the correct scaling of the
longitudinal size since $\xi_\parallel \sim L$).

These results strongly support a power-law behavior, i.e., $\tau\sim
L^z$.  In order to estimate $z$, we fit the data to the ansatzes $\tau
= a L^z$ [fitting tha data for $L\ge 16$, it gives $z=3.08(2)$ and
  $a=0.33(2)$] and $\tau= aL^3 + b L^2$ corresponding to $z=3$ with
the expected $O(1/L)$ corrections [it gives $a\approx 0.46(1)$ and
  $b=-0.7(2)$].  These fits work equally well as shown in
Fig.~\ref{tautc}.  We consider
\begin{equation}
  z=3.0(1)  \label{zestimate}
\end{equation}
as our final estimate (the central estimate $z=3$ is also supported by
the off-equilibrium simulations, see below).  The integrated
autocorrelation time
of the energy density gives substantially equivalent results.

Although this estimate of $z$ is obtained for heat-bath MC simulations
at $q=20$, we expect that it holds for any $q>4$.  Indeed, the value
of $z$ should be intrinsically related to the interface dynamics which
is expected to be shared by all $q>4$ at their FOTs. Moreover, it
should extend to the whole class of purely relaxational dynamics
(model A according to Ref.~\cite{HH-77}), including also the
Metropolis upgrading.

\subsection{Off-equilibrium scaling ansatzes}
\label{effsca}

A scaling theory for the off-equilibrium dynamics across $T_c$ can be
heuristically derived by scaling arguments, similar to those commonly
used at continuous transitions.

Assuming the existence of a nontrivial scaling behavior around $t=0$
corresponding to $T_c$, we may expect that it is controlled by two
scaling variables, such as
\begin{eqnarray}
r_1 = (t/t_s)^{-\nu}/L,\qquad  r_2 = t/L^z.
\label{s12}
\end{eqnarray}
In particular $r_1$ may be interpreted as the ratio between the
equilibrium correlation length at the given $\beta(t)$ and $L$.  The
off-equilibrium scaling behavior arising from  the protocol (\ref{betat})
is meant to describe the deviations of the statistical correlations from
the equilibrium scaling behavior,  when they cannot adapt themselves
to the changes of the temperature across $T_c$.
Thus equilibrium scaling should be recovered
in the limit $|r_2|\to\infty$ keeping $r_1$ fixed.

Equivalently we consider the scaling variables
\begin{eqnarray}
&u \equiv  {t_s^\kappa/L},\qquad
&\kappa = {\nu/(1 + z\nu)} ,
\label{udef}\\
&w \equiv  {t / t_s^{\kappa_t}},\qquad
&\kappa_t = z \kappa,
\label{wdef}
\end{eqnarray}
which are combinations of $r_1$ ad $r_2$.

We already note that the case of symmetric boundary conditions, such
as PBC, appears problematic, due to the divergence of the dynamic
exponent $z$.  In particular, taking the $z\to\infty$ limit of the
exponents $\kappa$ and $\kappa_t$ in the case of a square geometry,
one would naively obtain $\kappa=0$ and $\kappa_t=1$.  This may hint
at the absence of a nontrivial scaling behavior around $T_c$, i.e., we
may not observe a nontrivial off-equilibrium scaling behavior around
$t=0$ in the off-equilibrium protocol (\ref{betat}). Alternatively,
this may indicate a logarithmic scaling behavior, for example with
scaling variables $u\approx \ln(t_s)/L$ and $w \approx t/t_s$.  As we
shall see, numerical simulations favor a regular behavior around $t=0$
extending to $t/t_s>0$, which may be somehow related to a
metastability phenomenon.

Systems with MBC appear more promising to realize an off-equilibrium
scaling behavior. Indeed the corresponding values of the equilibrium
exponent $\nu$ and $z$ provide well defined exponents $\kappa$ and
$\kappa_t$ in Eqs.~(\ref{udef}) and (\ref{wdef}). In the case of the
FOT of the 2D Potts model, for which $\nu=1/2$ and $z=3.0(1)$, we
obtain $\kappa=0.200(4)$ and $\kappa_t=0.600(8)$, which lead to a
nontrivial power-law dependence of the scaling variables.  Therefore,
observing an off-equilibrium scaling behavior around $t=0$ is to be
expected in this case.

Our main working hypothesis is that the slow dynamics across $T_c$
presents a double scaling behavior in the large-$L$ limit, in terms of
the scaling variables $u$ and $w$, i.e.,
\begin{eqnarray}
&& m_r(t,t_s,L) \approx f_m(u,w) , \label{fm}\\
&& e_r(t,t_s,L) \approx f_e(u,w) . \label{fe}
\end{eqnarray}
We also expect that,  if we start 
from the high-$T$ phase [sign + in Eq.~(\ref{betat})],
the following asymptotic limits apply:
\begin{eqnarray}
&& \lim_{w\to -\infty} f_\#(u,w) = 0, \quad
\lim_{w\to \infty} f_\#(u,w) = 1,\label{lfm}
\quad
\end{eqnarray}
corresponding to
the large-$L$ equilibrium values at the two phases
(the subscript $\#$ corresponds to both $m$ and $e$).
The limits are reversed if we start from the low-$T$ phase 
[sign - in Eq.~(\ref{betat})].

Summarizing, in the cases where an off-equilibrium scaling behavior is
driven by slow variations across $T_c$, around $t=0$ corresponding to
$T(t)=T_c$, Eqs.~(\ref{fm}) and (\ref{fe}) with the scaling variables
(\ref{udef}) and (\ref{wdef}) are expected to provide the asymptotic
scaling behaviors, when $L$ is much larger than any other length
scale, thus when $L\gg \xi^\pm$ (see the end of Sec.~\ref{pottsmo}).
These asymptotic behaviors should be approached with power-law
$O(L^{-\omega})$ suppressed corrections, presumably $\omega=1$. Note
that for observables depending on the spatial coordinates one may add
the spatial scaling dependence on $x/L$.

In the case of MBC the off-equilibrium behavior is expected to be
related to the dynamics of the interface enforced by the MBC.  This
scenario implies a close relation between the scaling functions $f_m$
and $f_e$ of the magnetization and energy density. Let us assume that
the slowest modes, determining the off-equilibrium dynamics of the
magnetization and energy density, are associated with the wall
separating the spatial regions corresponding to the two different
phases.  If we define $x(t)$ as the location of the wall along the
$x_1$ axis, separating the low-$T$ from the high-$T$ region, we expect
that $m_r=e_r=1$ for $x_1<x(t)$ and $m_r=e_r=0$ for $x_1>x(t)$. Thus,
when the width of the domain wall becomes negligible, 
thus asymptotically for large $L$,
we expect that
\begin{equation}
m_r(t) \approx e_r(t) \approx {1\over 2} [1 + X_w(t)],
\label{mtet}
\end{equation}
where $X_w(t) = x(t)/L$ with $-1\le X_w(t) \le 1$.  As a trivial
consequence, we would have
\begin{equation}
f_e(u,w) = f_m(u,w). 
\label{femeq}
\end{equation}
Therefore, ${\cal X}_w(u,w) \equiv 2 f_m(u,w) -1$ may be considered as
an estimator of the average position of the interface in the large-$L$
limit.

The off-equilibrium scaling behaviors (\ref{udef}-\ref{fe}) at FOTs
are analogous to those expected at continuous transitions when slowly
crossing the critical temperature, usually related to the KZ
mechanism~\cite{Zurek-85,CEGS-12}.  The main difference is related to
the fact that at continuous transitions the critical exponents, such
as $\nu$ and $z$, do not depend on the boundary conditions (only the
scaling functions do).  For example, in the case of the 2D Ising model
corresponding to the $q=2$ Potts model, the off-equilibrium dynamics
of the magnetization (\ref{madef}) across the continuous transition is
expected to be
\begin{equation}
m(t,t_s,L) = L^{-\eta/2} f_m(u,w)
\label{mts}
\end{equation}
where $\eta=1/4$ and the scaling variables have the same form as those
in Eqs.~(\ref{udef}) and (\ref{wdef}). The exponents $\kappa$ and
$\kappa_t$ are obtained using the Ising exponents $\nu=1$ and
$z\approx 2.167$ associated with the purely relaxational dynamics
(see, e.g., Refs.~\cite{NB-00,CV-11} and references therein), i.e.,
$\kappa\approx 0.316$ and $\kappa_t\approx 0.684$.  At the Ising
critical point, different geometries and boundary conditions can only
change the scaling function $f_m$.

We finally mention that similar scaling arguments have been also
employed to study the effects of smooth spatial inhomogeneities at
FOTs~\cite{BDV-14,CNPV-15,WSK-15}, for example, when we assume a
spatially dependent temperature and look at the behavior of the system
around the spatial region where the temperature takes the value at the
transition point of the homogenous system.

\section{Monte Carlo simulations}
\label{MCsim}

In order to check the predictions of the previous section, we present
a numerical analysis of MC simulations following the off-equilibrium
protocol (\ref{betat}), for various geometries, boundary conditions,
lattice sizes and time scales. We use the heat-bath upgrading, see
App.~\ref{mehb}, varying the temperature according to
Eq.~(\ref{betat}), after each sweep which corresponds to a unit time.

\subsection{Square systems with PBC}
\label{respbc}

\begin{figure}[tbp]
\includegraphics*[scale=\graphicscale]{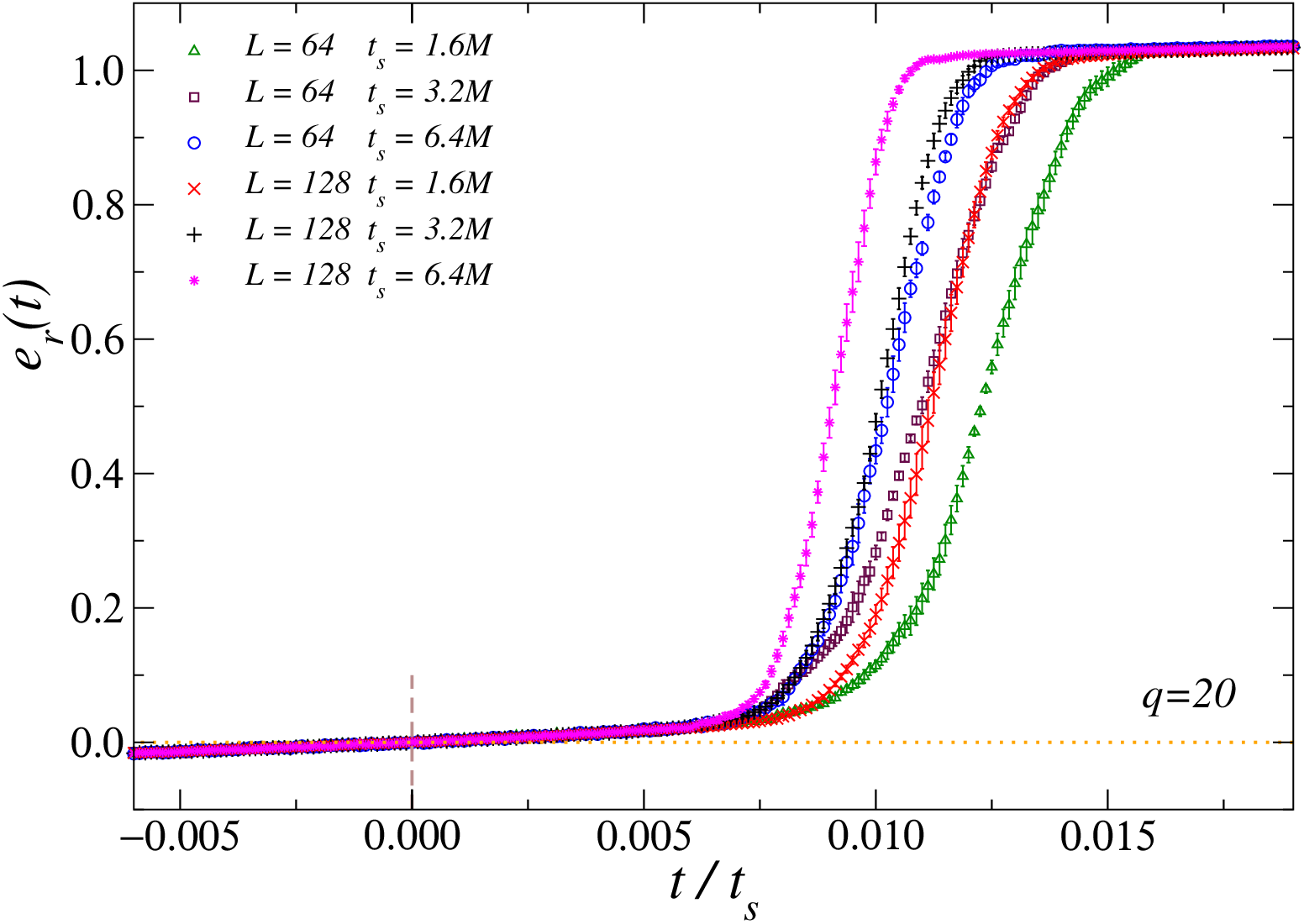}
\includegraphics*[scale=\graphicscale]{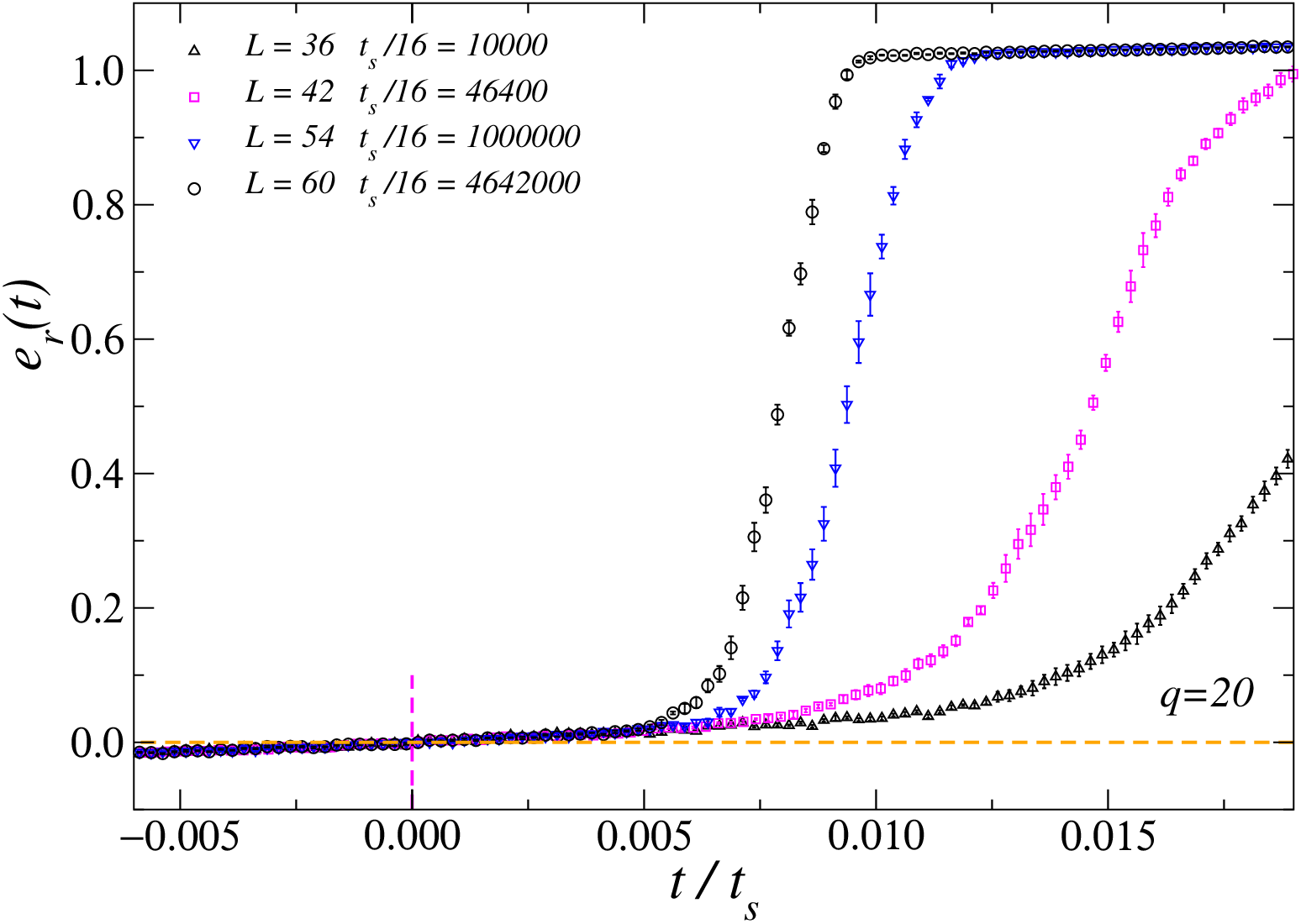}
%=% \vskip-5mm
\caption{(Color online) Data of $e_r(t)$ for the 2D $q=20$ Potts model
  in square $L^2$ lattices with PBC, starting from the high-$T$
  phase. We show two sets of data: (top) data for large lattices with
  $t_s=O(10^6)$; (bottom) data for smaller lattices, for which the
  ratio $\ln(t_s/16)/L\approx 0.2558$ is kept fixed (to check the
  possibility of a logarithmic scaling around $t=0$).  Both of them
  indicate a smooth behavior up to $t/t_s\approx 0.005$ and then a
  sharp relaxation toward the low-$T$ values, without showing
  nontrivial scaling behaviors around $t=0$.}
\label{plainpbcdata}
\end{figure}

To begin with, we present results for the square $L^2$ lattice with
PBC, and the off-equilibrium protocol (\ref{betat}) starting from the
high-$T$ phase.  In this case the magnetization vanishes by symmetry,
as a consequence of the average on the initial Gibbs ensemble at
$\beta(t_i)<0$.  Therefore we look at the behavior of the energy
density.  Figure~\ref{plainpbcdata} shows some results for the
renormalized energy density, cf. Eq.~(\ref{maener}), obtained by
heat-bath MC simulations of the $q=20$ Potts model for various $L$ and
time scales $t_s$.

When increasing $t_s$ and $L$, the data approach the corresponding
$L\to\infty$ equilibrium values up to $t=0$, where $e_r$ vanishes
within errors (we recall that $e_r=0$ is the $T\to T_c^+$ limit of the
equilibrium value in the thermodynamic limit). Then they increase
slowly up to $t/t_s=\tau^*>0$ with $\tau^*\approx 0.005$,
corresponding to $T(t)\approx 0.995\,T_c$, remaining well below the
equilibrium values of the low-$T$ phase. Then the data show a sharp
crossover to the values corresponding to the low-$T$ phase.  Note also
that the lower panel of Fig.~\ref{plainpbcdata} reports data with
$\ln(t_s)/L\approx {\rm const}$, which do not support scaling with
respect to the scaling variables $u\approx \ln(t_s)/L$ and $w \approx
t/t_s$.  Of course, we cannot exclude that an eventual logarithmic
scaling behavior may set in for larger $L$ and $t_s$.

The numerical results for medium-size $L=O(10^2)$ lattices and time
scales $t_s=O(10^6)$ suggest a smooth nonsingular behavior around
$t/t_s=0$, without hinting at nontrivial off-equilibrium scaling
behaviors around $t=0$.  This scenario was somehow anticipated in the
previous section, as a consequence of the trivial values of the
exponents $\kappa$ and $\kappa_t$ obtained using the scaling ansatzes
(\ref{udef}) and (\ref{wdef}).  It may be related to some form of
metastability developing in this slow cooling procedure, which likely
requires another theoretical framework.  This issue calls for further
investigation.

We expect that analogous scenarios occur at the FOTs of any $q>4$.

\subsection{Results for slab-like systems with MBC}

\begin{figure}[tbp]
\includegraphics*[scale=\graphicscale]{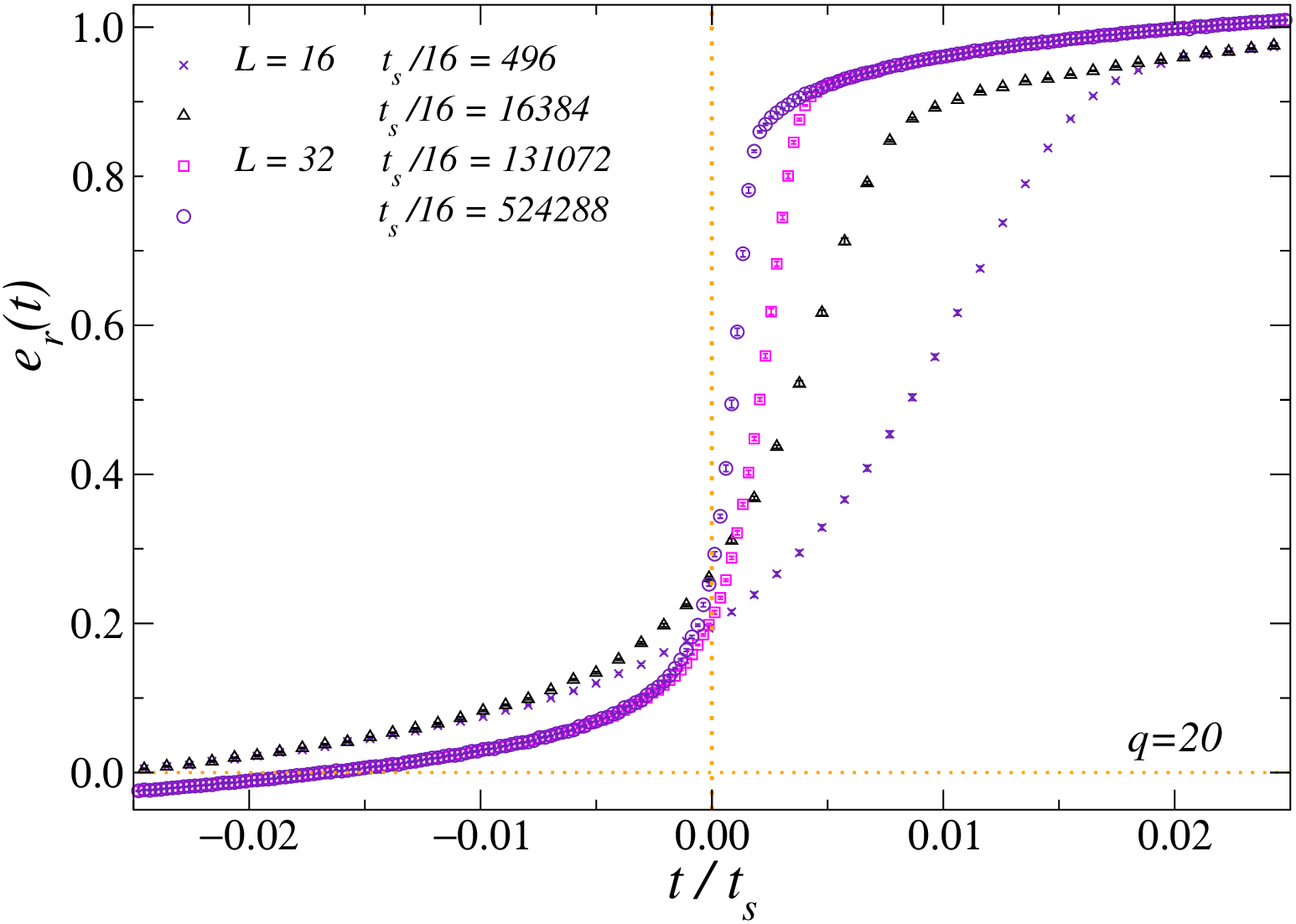}
\includegraphics*[scale=\graphicscale]{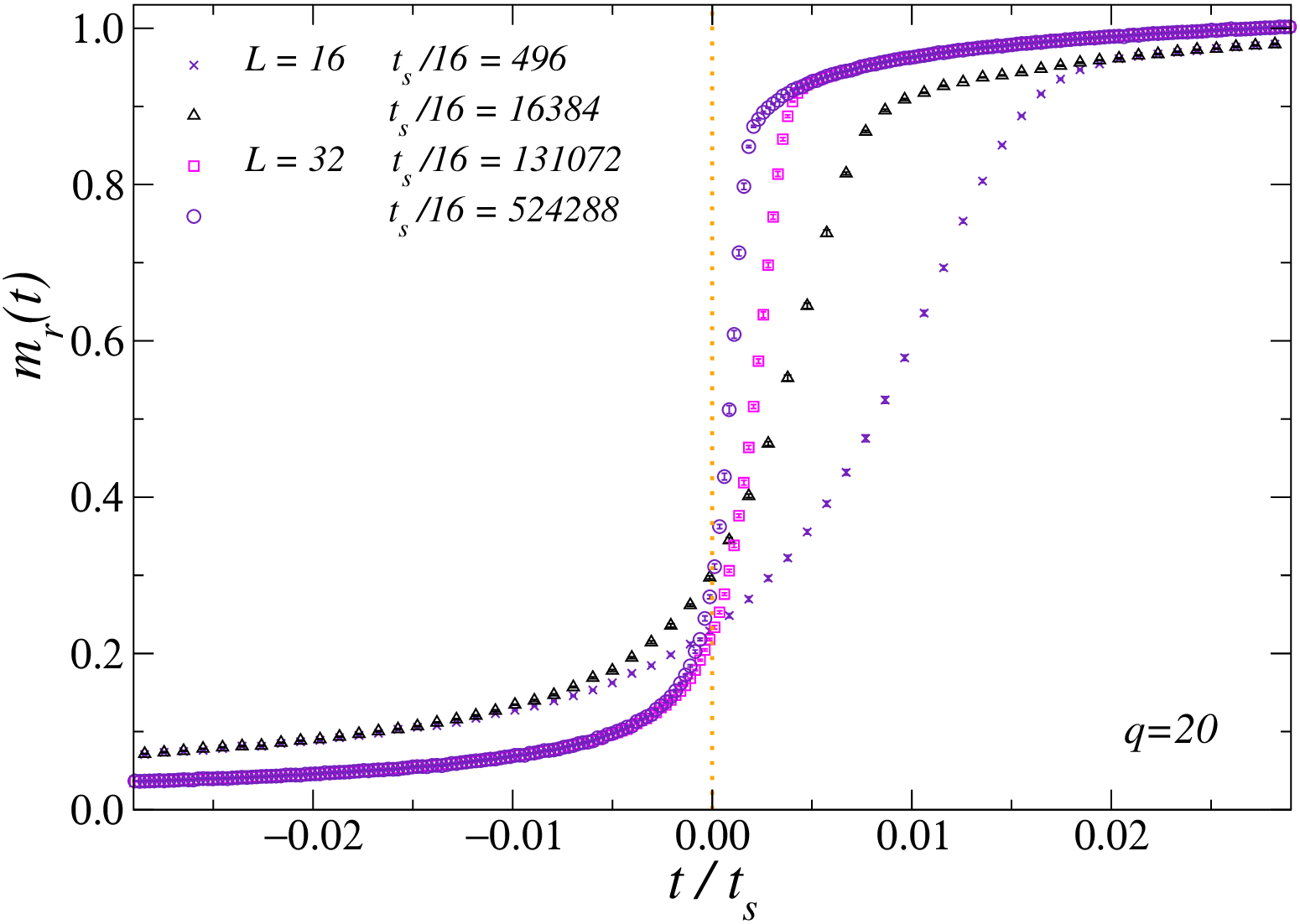}
%=% \vskip-5mm
\caption{(Color online) Data of $e_r(t)$ (top) and $m_r(t)$ (bottom)
  for the 2D $q=20$ Potts model with MBC, in anisotropic $L_\perp\times
  L_\parallel$ lattices with $L_\perp=2L+1$ and $L_\parallel=8L$, for
  $L=16$ and $L=32$, and various values of $t_s$.  }
\label{plaindata}
\end{figure}

\begin{figure}[tbp]
\includegraphics*[scale=\graphicscale]{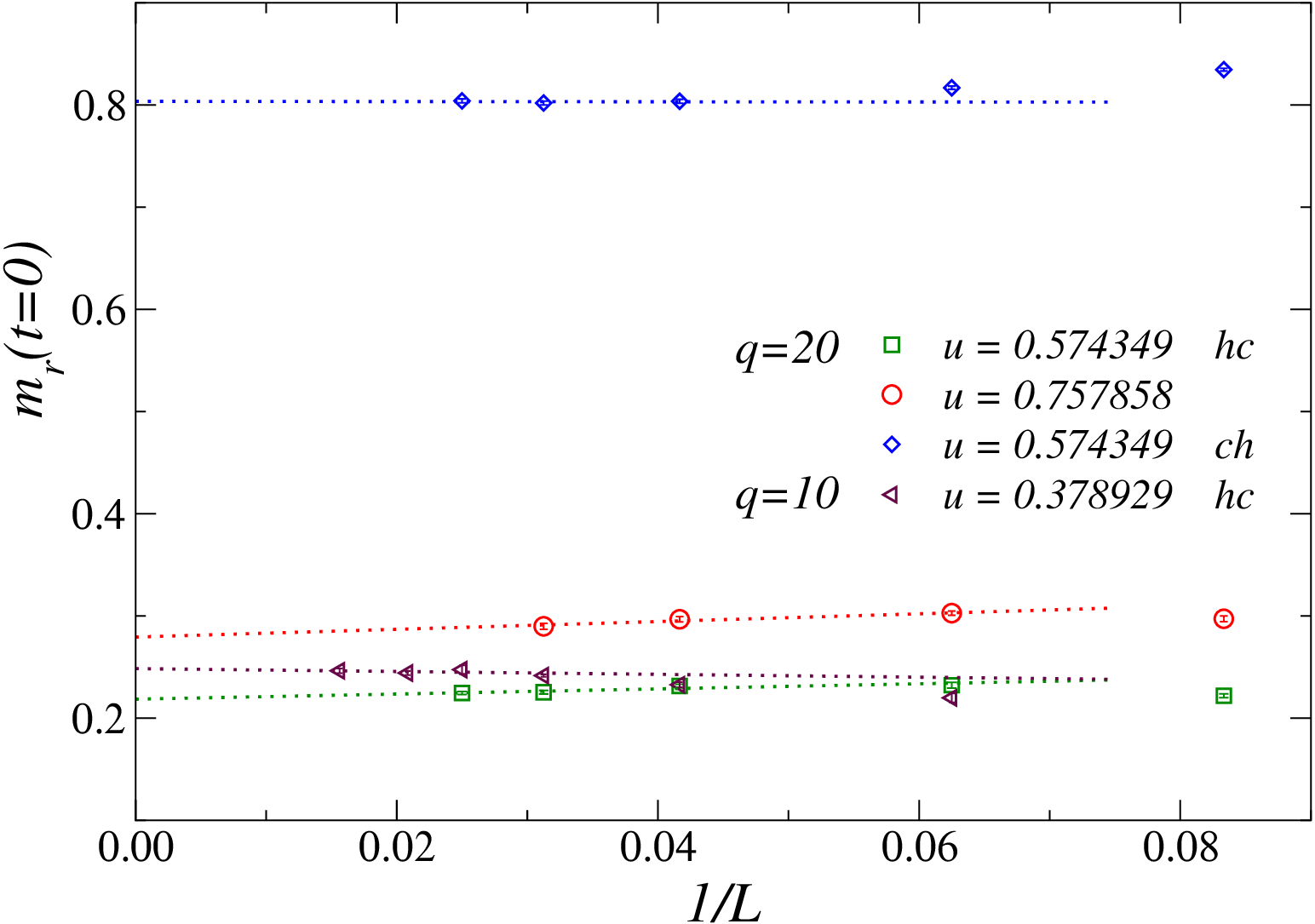}
%=% \vskip-5mm
\caption{(Color online) Data of $m_r(t=0)$ for $q=20$, starting from
  high-$T$ (hc) and low-$T$ (ch) phases, and for various values of
  $u\equiv t_s^{0.2}/L$ (which is kept fixed when varying $t_s$ and
  $L$) vs $1/L$.  They appear to approach a constant value with
  increasing $L$, supporting the general scaling ansatz (\ref{fm})
  with $\kappa=0.2$. The dotted lines show a linear fit of the data
  for the largest available lattice sizes.}
\label{datat0}
\end{figure}

We now present results for the 2D Potts models with $q=10$ and $q=20$,
in anisotropic $(2L+1)\times L_\parallel$ lattices with MBC, starting
from the high-$T$ and low-$T$ phase.  We consider the slab-limit
$L_\parallel \gg L_\perp$, for which numerical results can be
straightforwardly obtained by increasing the longitudinal size up to
the point where the data appear stable within the errors.  We checked
that $L_\parallel=8L$ turns out to be sufficiently large to
effectively provide infinite-$L_\parallel$ results within the errors
(the linear scaling of $L_\parallel$ with $L$ takes into account that
$\xi_\parallel\sim L_\perp$ for MBC).  In our numerical simulations we
choose $|\delta(t_i)|=|\delta(t_f)|=1/32$. However, as we shall see,
the emerging scaling behavior does not depend on these particular
values, because it is essentially related to the behavior around $t=0$
where $T(t)=T_c$.

In Fig.~\ref{plaindata} we report some raw data during a slow
variation of the temperature across $T_c$ for the slab geometry with
MBC.  They are obtained starting from the high-$T$ phase and show the
behavior of the energy and the magnetization when the transition is
crossed with different time scales $t_s$ and lattice sizes $L$.  The
data show that the effective passage from one phase to the other
occurs around $t/t_s=0$. As we shall see, the dependence of the
various curves on $L$ and $t_s$ can be cast in the off-equilibrium
scaling behavior put forward in the previous section,
cf. Eqs.~(\ref{udef}-\ref{fe}).

\begin{figure}[tbp]
\includegraphics*[scale=\graphicscale]{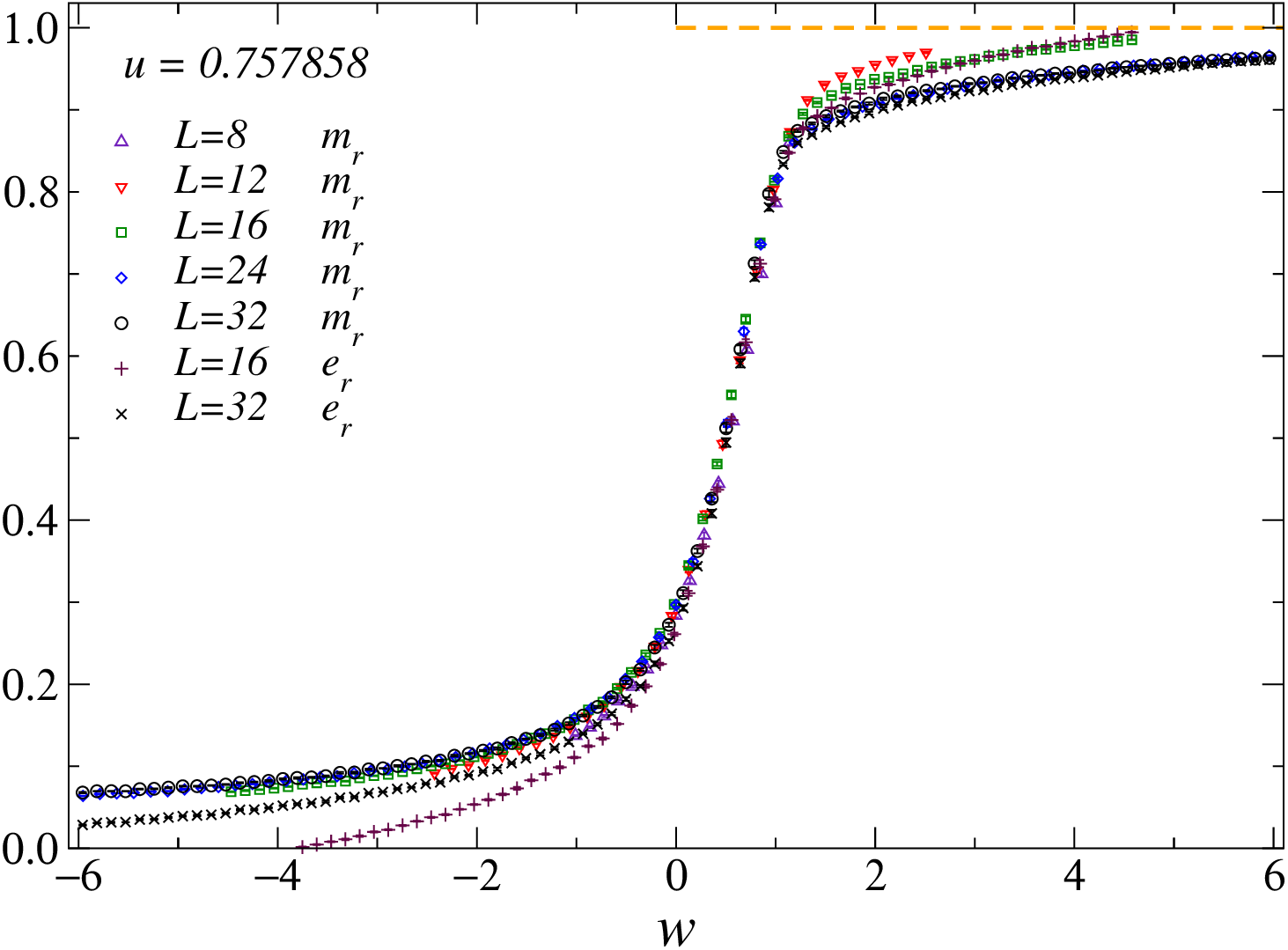}
\includegraphics*[scale=\graphicscale]{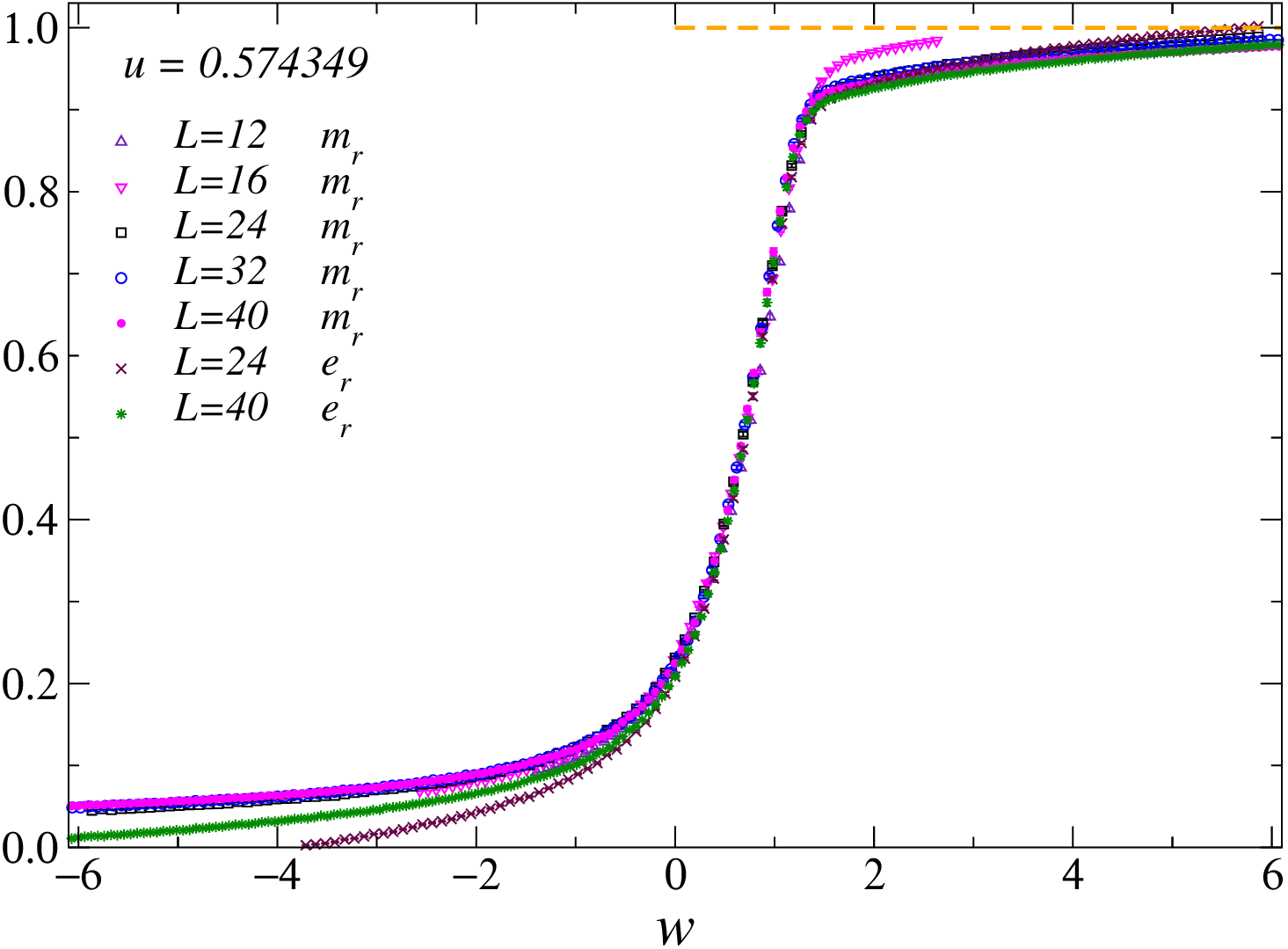}
%=% \vskip-5mm
\caption{(Color online) Data of $m_r(t)$ and $e_r(t)$ for $q=20$
  starting from the high-$T$ phase, i.e., $T>T_c$, for two values of
  $u=t_s^{\kappa}/L$ with $\kappa=0.2$, i.e., $u\approx 0.7578$ (top)
  and $u\approx 0.5743$ (bottom).  With increasing $L$, they approach
  asymptotic curves when plotted versus $w=t/t_s^{\kappa_t}$ with
  $\kappa_t=0.6$, supporting the scaling ansatzes (\ref{fm}) and
  (\ref{fe}). Moreover, the data are consistent with the relation
  (\ref{femeq}) predicting the same asymptotic curve for $m_r$ and
  $e_r$.  }
\label{datascal}
\end{figure}

\begin{figure}[tbp]
\includegraphics*[scale=\graphicscale]{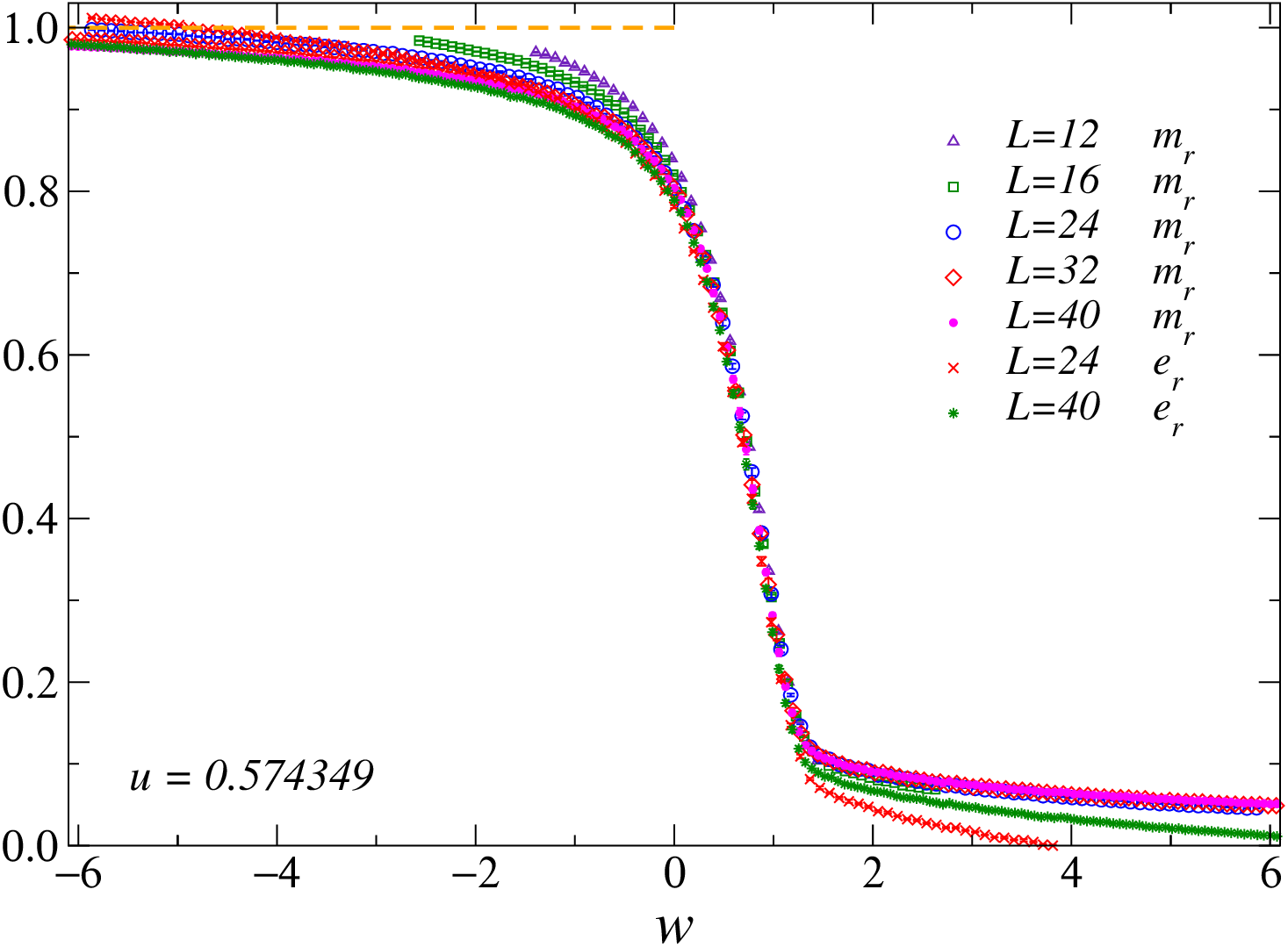}
%=% \vskip-5mm
\caption{(Color online) Data for $m_r(t)$ and $e_r(t)$ starting from
  the cold phase, i.e., $T<T_c$, keeping the scaling variable
  $u\approx 0.5743$ fixed.  They approach a unique asymptotic curve
  with increasing $L$, when plotted versus $w=t/t_s^{\kappa_t}$,
  supporting the scaling ansatzes (\ref{fm}) and (\ref{fe}), and the
  interface relation (\ref{femeq}).  }
\label{datascalch}
\end{figure}

\begin{figure}[tbp]
\includegraphics*[scale=\graphicscale]{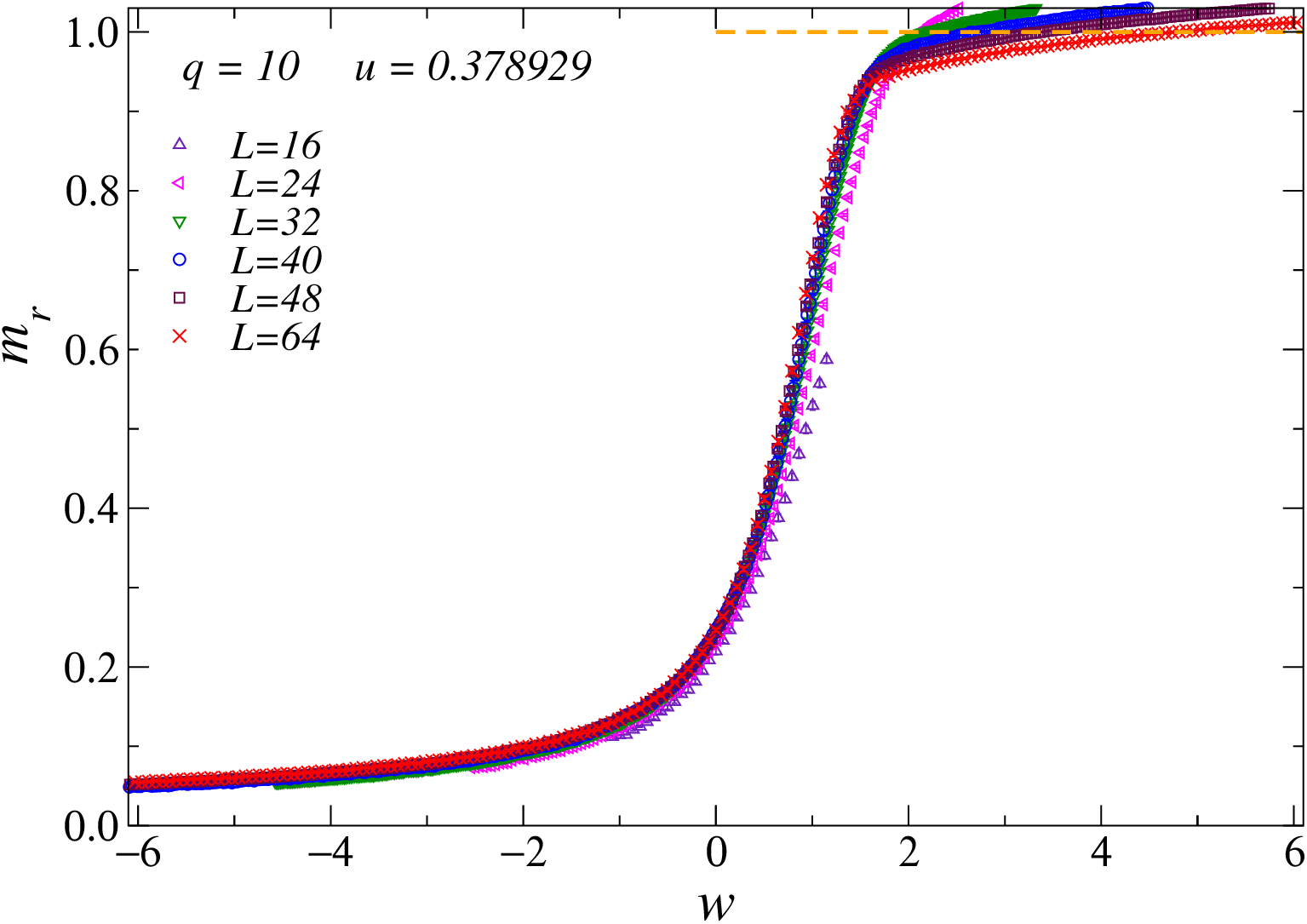}
%=% \vskip-5mm
\caption{(Color online) Data of $m_r(t)$ for $q=10$ versus
  $w=t/t_s^{\kappa_t}$ at a fixed value of $u=t_s^{\kappa}/L\approx
  0.3789$ (with $\kappa_t=0.6$ and $\kappa=0.2$), starting
from the high-$T$ phase.  They approach an
  asymptotic scaling curve, in agreement with Eq.~(\ref{fm}).  }
\label{dataq10}
\end{figure}

In order to check the scaling in the variable $u=t_s^\kappa/L$, we
first note that Eqs.~(\ref{fm}) and (\ref{fe}) imply that at $t=0$
\begin{equation}
m_r(0,t_s,L) \approx g_m(u),\qquad
e_r(0,t_s,L) \approx g_e(u).
\label{mrer0}
\end{equation}
Therefore, we expect that data at $t=0$ and fixed $u=t_s^\kappa/L$
must converge to nontrivial $u$-dependent values with increasing $L$.
Fig.~\ref{datat0} shows data at some fixed values of $u$ (using
$\kappa=0.2$ obtained taking the central value $z=3$), for $q=20$ and
$q=10$. They appear to converge to nontrivial values, supporting the
above asymptotic behavior, with corrections which approximately decay
as $O(L^{-1})$.

Analogous results are obtained by slightly changing the value of $z$,
according to the equilibrium estimate $z=3.0(1)$, corresponding to
$\kappa = 0.200(4)$. Actually, one may assume the off-equilibrium
scaling (\ref{mrer0}) to independently estimate $z$ from the
off-equilibrium data. For example, by allowing for deviations from
$z=3$, i.e., $z=3+\delta z$, and interpreting the scaling corrections
of the data of Fig.~\ref{datat0} as due to $\delta z$, we find again
that the optimal value is $z=3$ with a few percent of uncertainty
(corrections to scaling have different sign in some cases, which
cannot be explained by a unique shift of $z$, thus $z\approx 3$
appears as the optimal value).

Note also that, since the equilibrium average position of the
interface at $T_c$ is expected at equal distances from the boundaries
in the large-$L$ limit, leading to the asymptotic equilibrium values
$m_r(T_c) = e_r(T_c) = 1/2$, we expect that $\lim_{u\to \infty}
g_\#(u) = 1/2$.

The scaling with respect to $w$ at fixed values of $u$ is supported by
the plots in Figs.~\ref{datascal}, \ref{datascalch}, and
\ref{dataq10}, respectively for $q=20$ with hot and cold starting
point and $q=10$ with hot start.  In all cases the data approach an
asymptotic function of the scaling variable $w$, as predicted by the
scaling theory of Sec.~\ref{offsca}.  Moreover, the data of $m_r(t)$
and $e_r(t)$ shown in Figs.~\ref{datascal} and \ref{datascalch}
approach the same scaling curve, in agreement with the asymptotic
relation (\ref{femeq}), thus fully supporting the hypothesis that the
off-equilibrium behavior is essentially controlled by the
time-dependent position of the interface.

\begin{figure}[tbp]
\includegraphics*[scale=\graphicscale]{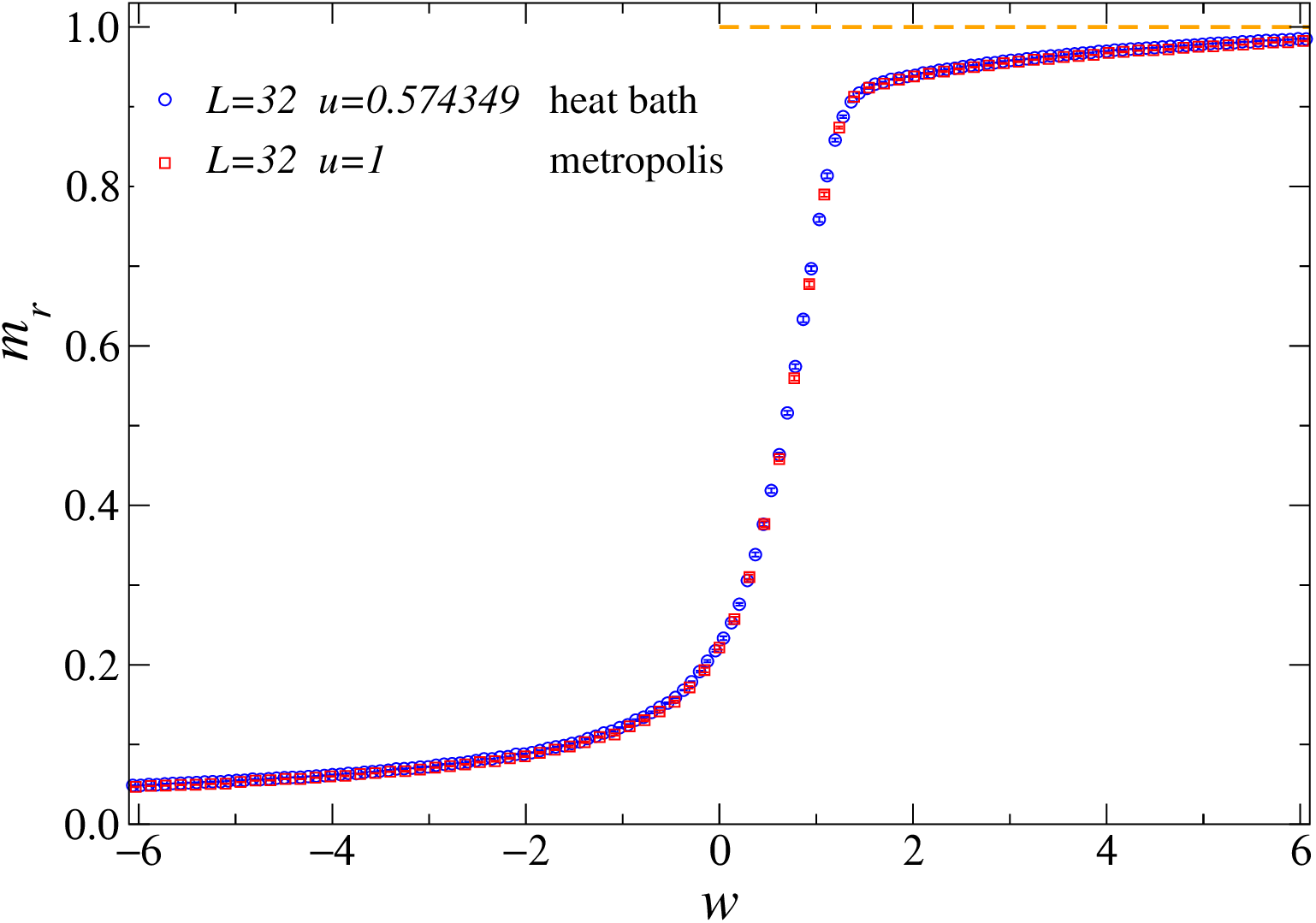}
%=% \vskip-5mm
\caption{(Color online) Check of universality between the heat-bath
  and Metropolis dynamics.  The scaling curves coincide after a
  rescaling of the scaling variables $u$ and $w$ of the Metropolis
  data.  In particular, the rescaling of the variable $u$ can be fixed
  by looking for data with equal $m_r(0)$ (we obtain that $u\approx
  0.5743$ of the heat-bath dynamics approximately corresponds to $u=1$
  of the Metropolis dynamics).}
\label{datamh}
\end{figure}

Finally, in Fig.~\ref{datamh} we compare results obtained using the
heat-bath algorithm with those obtained by a standard Metropolis
algorithm with only one trial per site, see App.~\ref{mehb}.  The
curves match after a trivial rescaling of the scaling variables $u$
and $w$, supporting the expected universality with respect to the type
of relaxational dynamics.

We conclude that our numerical analysis for slab geometries with MBC
supports the off-equilibrium scaling behaviors of the magnetization
and energy density put forward in Sec.~\ref{effsca},
cf. Eqs~(\ref{udef}-\ref{femeq}).

\section{Conclusions}
\label{conclu}

We investigate off-equilibrium behaviors at FOTs driven by a time
dependence of the temperature across the transition point $T_c$.
Usually, off-equilibrium behaviors at FOTs are associated with
phenomena of metastability and hysteresis~\cite{Binder-87}.  We focus
on the possibility of nontrivial off-equilibrium scaling behaviors
driven by slow changes of the temperature, analogous to those arising
at continuous transitions, leading to the KZ
mechanism~\cite{Kibble-76,Zurek-85}.  When slowly varying the
temperature across a continuous transition, for example linearly as
$T(t)/T_c\approx 1 - t/t_s$, the KZ mechanism predicts a nontrivial
off-equilibrium scaling behavior in the limit of slow variations, due
to the diverging length scale at $T_c$.  We show that phenomena
analogous to the KZ off-equilibrium scaling emerge also at FOTs, when
appropriate boundary conditions are considered.

We consider the 2D Potts models, which provide an ideal testing ground
to investigate issues related to FOTs. In our discussion we consider a
purely relaxational dynamics such as that obtained by heat-bath and
Metropolis upgrading in MC simulations. We study the off-equilibrium
behavior in the case of a time-dependent temperature crossing the FOT.
In particular, we consider a linear dependence of the inverse
temperature $\beta(t) = \beta_c (1 \pm t/t_s)$, starting from the
high-$T$ or low-$T$ phase.

We point out that off-equilibrium behaviors at FOTs are extremely
sensitive to the geometry and boundary conditions of the system.  This
peculiar dependence is essentially related to the equilibrium
relaxational dynamics at $T_c$. For symmetric boundary conditions,
such as PBC, we expect an exponentially slow dynamics due to an
exponentially large tunneling time $\tau\sim e^{\sigma L}$. On the
other hand, a power-law slowing down $\tau\sim L^z$ with $z\approx
3.0$ is found when considering MBC, i.e., when the boundary conditions
at two opposite sides of the system are related to the different
high-$T$ and low-$T$ phases, effectively generating an interface
separating the coexisting phases.  We argue that an off-equilibrium
scaling behavior around $t=0$ (i.e., the time corresponding to $T_c$)
is realized for MBC. This is controlled by the corresponding
equilibrium length-scale and dynamic exponents. We argue that this
scaling behavior is essentially related to the dynamics of the
interface enforced by MBC.

In the case of slab-like geometries with MBC, the numerical results
for the $q=10$ and $q=20$ Potts model support the emergence of an
off-equilibrium scaling picture characterized by power-law behaviors,
analogous to those predicted by the KZ theory at continuous
transitions.

On the other hand, symmetric boundary conditions do not apparently
lead to nonanalytic scaling behaviors at $t=0$, but rather to a
delayed sharp relaxation to the other phase at $t/t_s\gtrsim \tau^*$
with $\tau^*>0$, which may be related to a metastability phenomenon
arising from the slow dynamics across the FOT.  This point deserves
further investigation to physically understand it. Further checks of
the observed asymptotic behavior may also be called for.  Indeed, we
cannot exclude the possibility that a different (logarithmic)
asymptotic behavior sets in for sizes and time scales larger than
those considered in our numerical analysis, which are $L=O(10^2)$ and
$t_s=O(10^6)$.

Our scaling arguments are quite general, therefore they should also
apply to higher-dimensional systems, such as the 3D Potts models that
undergo FOTs.  Such an extension may generally depend on the geometry
of the system, e.g., cubic-like $L^3$, slab-like $L_\perp\times
L_\parallel^{d-1}$ and tube-like $L_\perp^2 \times L_\parallel$ (with
$L_\parallel\gg L_\perp)$ geometries, as well as on the boundary
conditions. In particular, we again expect that geometries and
boundary conditions favoring the emergence of an interface should give
rise to off-critical scaling behaviors similar to those of the KZ
theory at continuous transitions.  This issue calls for further
investigation.

Off-equilibrium scaling behaviors may also appear at FOTs driven by
magnetic fields.  For example, one may consider O($N$)-symmetric spin
models in the low-$T$ ordered phase, where FOTs are driven by an
external magnetic field coupled to the spin variables. Then, one may
consider the off-equilibrium dynamics driven by a time-dependent
magnetic field $h(t) = t/t_s$ across the transition point $h=0$.  We
expect that in the case of Ising models ($N=1$) an off-equilibrium
scaling behavior may emerge in systems with boundary conditions
enforcing the presence of an interface, analogously to what is
observed at the thermal FOTs of the 2D Potts models.  In the case of a
continuous symmetry ($N>2$), off-equilibrium scaling behavior may
emerge from the spin-wave dynamics (Goldstone modes related to the
broken O($N$) symmetry)~\cite{FP-85}.  We believe that these issues
are worth being further investigated.

One may also consider evolutions different from the purely relaxational
ones. They would generally lead to other values of the dynamic
exponent~$z$.

We also mention that off-equilibrium behaviors arising from sudden
quenches below and at the transition point have been discussed in
several works, see,
e.g.,~\cite{GSS-83,Bray-94,SZ-00,OKIM-03,GPW-04,LJ-07,PI-07}.

Off-equilibrium behaviors at FOTs are quite general, they should be
observable in many physical contexts where the FOTs are approached by
varying the system parameters. The off-equilibrium protocol
investigated in this paper may be exploited to probe the main features
of systems at the FOT. Moreover, our results may turn out useful in
understanding more complicated off-equilibrium phenomena at FOTs. For
example, as a case of physical interest we mention the effects of the
intrinsic space-time inhomogeneity of the quark-gluon plasma formation
in heavy-ion collisions~\cite{qgp-ref}, whose equilibrium $T$-$\mu$
($\mu$ is the chemical potential) phase diagram is expected to have a
FOT line~\cite{RW-00} which may be crossed during heavy-ion
collisions. Another interesting context is that of the universe
cosmology, which was the original ground of the Kibble
proposal~\cite{Kibble-76} to understand the effects of an expanding
universe through a continuous transition.  From analogous studies of
off-equilibrium behaviors at FOTs we may learn the effects of an
expanding and cooling universe that passes through a FOT.

Analogous off-equilibrium phenomena should be also observable in
quantum many-body systems, at first-order quantum transitions.  Some
issues arising from slow (quasi-adiabatic) passages through quantum
FOTs have been recently discussed, in particular for some
one-dimensional quantum chains~\cite{CPV-15,SD-15,YKS-10}, including
issues related to adiabatic evolutions in quantum
computations~\cite{CPV-15,AC-09,MRB-09,LMSS-12}.

\bigskip

{\bf Acknowledgements:}
H.P. would like to acknowledge INFN, Sezione di Pisa, for the kind
hospitality. H.P. acknowledges partial support from the Research
Promotion Foundation of Cyprus, under grant
TECHNOLOGY/$\Theta$E$\Pi$I$\Sigma$/0311(BE)/16.

\bigskip

\appendix

\section{Metropolis and Heat-bath updatings of the Potts models}
\label{mehb}

A heat-bath updating of a single site variable
consists in the change $s_{\bf x} \to s_{\bf x}'$
with probability $\sim e^{-H(s_{\bf x}')/T}$ independent of the
original spin $s_{\bf x}$.  

The Metropolis updating of a single spin $s_{\bf x}$ is performed by
(i) proposing a new spin $s_{\bf x}'\ne s_{\bf x}$ by taking one of
the other $q-1$ states with equal probability, (ii) accepting the
change with probability ${\rm Min}[e^{[H(s_{\bf x})-H(s_{\bf
        x}')]/T},1]$.

The heat-bath updating is generally more effective than a single
Metropolis updating, because the new variable is not correlated with
the previous one. The Metropolis updating tends to be equivalent to
the heat-bath one when a large number of trials are performed.

Both updating procedures give rise to a purely relaxational dynamics,
usually named model A~\cite{HH-77}, whose class also includes
configuration updatings by Langevin equations with white noise.
The time unit during the relaxational dynamics is generally
associated with a complete sweep of the lattice variables.

\section{Computation of the equilibrium autocorrelation time}
\label{autoco}

The integrated autocorrelation time of a given quantity $Q$ is defined
as
\begin{eqnarray}
\tau \equiv {1\over 2} \sum_{t=-\infty}^{t=+\infty} {C(t)\over C(0)},
\label{taudef}
\end{eqnarray}
where $C(t) = \langle \left( Q(t) - \langle Q \rangle \right) \left(
Q(0) - \langle Q \rangle \right) \rangle$ is the autocorrelation
function of $Q$ ($t$ is the discrete Monte Carlo time, where a time
unit is given by a sweep, i.e.,  a heat-bath update of all lattice
variables).  Averages are taken at equilibrium.  Estimates of the
corresponding integrated autocorrelation time $\tau$ can be obtained
by the binning method (see, e.g., Ref.~\cite{Wolff-04,DMV-04} for
discussions of this method and its systematic errors), using the
estimator
\begin{equation}
\tau =  {E^2\over 2 E_0^2},
\label{eqbl}
\end{equation}
where $E_0$ is the naive error calculated without taking into account
the autocorrelations, and $E$ is the correct error found after
binning, i.e., when the error estimate becomes stable with respect to
increasing of the block size $b$.  The statistical error $\Delta\tau$
is just given by $\Delta \tau/ \tau = \sqrt{2/n_{b}}$ where $n_{b}$ is
the number of blocks corresponding to the estimate of $E$. As
discussed in Ref.~\cite{Wolff-04} this procedure leads to a systematic
error of $O(\tau/b)$.  In our cases the ratio $\tau/b$ will always be
much smaller than the statistical error, so we will neglect it.
Eq.~(\ref{eqbl}) can be easily extended to the case the quantity $Q$
is measured every $n_m$ sweeps, i.e., $\tau = n_m E^2/(2 E_0^2)$, which
is of course meaningful only if $n_m\ll \tau$.

Of course, $\tau$ depends on the quantity $Q$ considered. However,
barring unlikely exceptions, all quantities lead to the same
asymptotic power-law behavior $\tau\sim L^z$.

\end{document}